\documentclass[aps,prl,twocolumn,superscriptaddress]{revtex4-2}

\usepackage{amsmath,amssymb}
\usepackage{hyperref}
\usepackage{physics}
\usepackage{graphicx}
\usepackage{color}
\begin{document}

\title{Coulomb blockade in a {non-thermalized} quantum dot}

\author{G. McArdle}
\affiliation{School of Physics and Astronomy, University of Birmingham, Birmingham, B15 2TT}

\author{R. Davies}
\affiliation{School of Computer Science and Digital Technologies, Aston University, Birmingham, B4 7ET}

\author{I. V. Lerner}
\affiliation{School of Physics and Astronomy, University of Birmingham, Birmingham, B15 2TT}

\author{I. V. Yurkevich}
\affiliation{School of Computer Science and Digital Technologies, Aston University, Birmingham, B4 7ET}


\date{\today}

\begin{abstract}
We investigate non-equilibrium transport properties of a quantum dot in the Coulomb blockade regime under the condition of negligible inelastic scattering during the dwelling time of the electrons in the dot. Using the quantum kinetic equation we show that the absence of thermalization leads to a double-step in the distribution function of electrons on the dot, provided that it is symmetrically coupled to the leads. This drastically changes nonlinear transport through the dot resulting in an additional (compared to the {thermalized} case) jump in the conductance at voltages close to the charging energy, which could serve as an experimental manifestation of the absence of thermalization.
\end{abstract}

\maketitle

Many-body localization (MBL), predicted for disordered many-electron systems which are not thermalized with the environment  \cite{BAA2006, MBL2}, has attracted a lot of theoretical and experimental attention (for a review see \cite{MBLReviewPapic}) and has been observed in systems of ultracold atoms \cite{MBLExpt}.  One of the defining properties of MBL is the absence of thermalization  \cite{NandkishoreHuse, ScarsReview}.

Prior to the MBL papers \cite{BAA2006,MBL2}, a similar regime of localization in Fock space  {was} {predicted for quantum dots} \cite{AKGL} where electrons fail to mutually equilibrate as their dwelling time on the dot, $\tau_{\mathrm{dw}}$, is much shorter  {than} the equilibration time $\tau_{\mathrm{eq}}$.  {Alternatively, this condition can be formulated as}
\begin{align}\label{g<G}
    \gamma \ll \Gamma,
\end{align}
where  $\gamma \sim1/ \tau_{\mathrm{eq}}$ is the equilibration rate and {$\Gamma\sim 1/\tau_{\mathrm{dw}}$} is the tunneling rate.
For  a zero-dimensional diffusive dot, the electron-electron equilibration rate \cite{AKGL, SivanImry1994, BlanterRates},
\begin{equation}\label{QP Rate}
	\gamma \approx \Delta \left(\frac{\varepsilon}{g\Delta}\right)^2\!,
\end{equation}
can be sufficiently small provided that  {$\sqrt{g}\Delta < \varepsilon < g\Delta$}, {where $\varepsilon $ is the quasiparticle energy,} $\Delta$ is the mean level spacing on the dot, and $g\Delta $ is the Thouless energy of the dot with dimensionless conductance $g \gg 1$.

In this Letter, we show that such an absence of thermalization leads to striking changes in nonlinear transport in the Coulomb blockade regime, where electrons are loaded one-by-one into a quantum dot due to the charging energy, $E_\mathrm{c}=e^2/C$, of a dot of capacitance $C$, preventing a continuous flow.  We assume the separation of scales typical for the classical Coulomb blockade at a temperature $T$ (see \cite{Kouwenhoven1997, AleinerReview, AlhassidReview} for reviews):
\begin{equation}\label{scales}
	\Gamma \ll \Delta \ll T \ll E_\mathrm{c}.
\end{equation}
Typically, the study of quantum dots in the Coulomb blockade regime has been focused on the regime where complete thermalization is assumed. This regime is characterized by peaks in the conductance as a function of gate voltage \cite{AvLikPaper,Beenakker} that can be attributed to interesting features in the {tunneling} density of states \cite{TDoS}, and -- in case of strong asymmetry in the coupling to the leads -- by a staircase in the current as a function of the bias voltage $V$ \cite{Kulik,AverinLikharevBookChapter, Ben-JacobWilkins, AvLikKor}. When the coupling is approximately symmetric, $\Gamma_\mathrm{L} \sim \Gamma_\mathrm{R}$, the Coulomb staircase practically vanishes in the thermalized case. But it is precisely in this case when the absence of thermalization reveals itself by an additional jump in the nonlinear differential conductance, as we show in this Letter by solving the  quantum kinetic equation. {The absence of thermalization on a dot, therefore,} can be detected by this jump {which occurs within the first step of the Coulomb staircase.}

 The jump arises due to the change in the distribution function of the dot; going from a Fermi function in the fully {thermalized} case to a double-step form. A similar structure (although for practically noninteracting electrons) has previously been observed in one-dimensional wires where the distribution function was a linear combination of the two Fermi functions of the leads due to insufficient time for equilibration \cite{Birge}. A double-step distribution has also been predicted for open quantum dots, where electrons are practically noninteracting \cite{AltlandEgger}, and for auxiliary non-interacting electrons in the slave-boson approach to the Kondo effect in quantum dots \cite{SmirnovKondoQD}. Here, in the Coulomb-blockade regime in region (\ref{scales}), a double-step form of the electron distribution function is substantially modified by the interaction.

 The standard Hamiltonian of a Coulomb-blockaded quantum dot coupled to two leads is  $H = H_\mathrm{dot} + H_\mathrm{l} + H_\mathrm{tun}$, where
\begin{subequations}\label{Hamiltonian}
\begin{align}
    H_\mathrm{dot} &= \sum_n \varepsilon_n d_n^\dagger d_n + \tfrac{1}{2} E_\mathrm{c} \big(\hat{N} - N_\mathrm{g} \big)^2, \\
	H_\mathrm{l} &= \sum_{k, \alpha} (\varepsilon_k-\mu_\alpha) c_{k, \alpha}^\dagger c_{k,\alpha}, \\
	H_\mathrm{tun} &= \sum_{k, n, \alpha} \big(t_{  \alpha } c_{k,\alpha}^\dagger d_n + \mathrm{h.c.} \big).
\end{align}
\end{subequations}
Here $\alpha {=} \mathrm{L, R}$ labels the leads, $d_n^\dagger (d_n), c^{\dagger}_{k, \alpha} (c_{k, \alpha})$ are the creation (annihilation) operators for electrons with energies $\varepsilon_n$ and $\varepsilon_k$ in the dot and leads respectively, $\hat{N} = \sum_n d_n^\dagger d_n$ is the number operator for the dot, {and} $N_\mathrm{g}$ is the preferred number of electrons on the dot set by the gate voltage. The leads have chemical potentials $\mu_\mathrm{L} = \mu+eV$ and $\mu_\mathrm{R} = \mu$. The  {$k$- and $n$-independent} tunneling amplitudes between the dot and leads, $t_\alpha$,   {define, along with the density of states of the leads $\nu_\alpha$ (taken to be constant),  the tunneling} rates $\Gamma_\alpha = 2 \pi \nu_\alpha |t_\alpha|^2$ with the total $\Gamma=\Gamma_{\mathrm{L}}+\Gamma_{\mathrm{R}}$.

In addition to inequalities (\ref{scales}), we assume that the Fermi energy of the dot is much larger than the charging energy, $\varepsilon_\mathrm{F} \gg E_\mathrm{c}$, to ensure that only electrons in a relatively narrow energy strip around $\varepsilon_\mathrm{F}$ contribute  the transport properties of the system. This assumption is also utilized in the orthodox theory of the Coulomb blockade {\cite{AverinLikharevBookChapter, Beenakker, Kulik, Ben-JacobWilkins, AvLikKor}} and is achievable in experiments \cite{Kouwenhoven1997, StaircaseExpt}. By starting with the standard expression for tunneling current \cite{MeirWingreenJauho} and assuming current conservation, we express the current across a quantum dot in the Coulomb blockade regime in the region \eqref{scales}  as
\begin{widetext}
\begin{equation}\label{Current}
	I = e\frac{\Gamma_\mathrm{L} \Gamma_\mathrm{R}}{\Gamma} \sum_{N,n} p_N \Big( F_N(\varepsilon_n)\left[f_\mathrm{L}(\varepsilon_n +\Omega_{N-1}) - f_\mathrm{R}(\varepsilon_n +\Omega_{N-1})\right] + (1-F_N(\varepsilon_n)) \left[f_\mathrm{L}(\varepsilon_n +\Omega_N) - f_\mathrm{R}(\varepsilon_n +\Omega_N) \right] \Big),
\end{equation}
\end{widetext}
with details of the derivation in \emph{Supplemental Material}.
Here $p_N$ is the probability of $N$ electrons being on the dot, $F_N(\varepsilon_n)$ is their distribution function, and $f_\mathrm{L,R}(\varepsilon_n)$ are Fermi functions in the leads with chemical potentials $\mu_\mathrm{L} = \mu+eV$ and $\mu_\mathrm{R} = \mu =\varepsilon_\mathrm{F} $ respectively. The presence of the charging energy is encapsulated by
\begin{equation}\label{OmegaN}
	\Omega_N = E_{N+1}-E_N = E_{\mathrm{c}}\left(N+\tfrac{1}{2}-N_{\mathrm{g}}\right),
\end{equation}
where $E_N = \frac{1}{2}E_\mathrm{c}(N-N_\mathrm{g})^2$.

The current through a {thermalized} quantum dot is usually considered with the {help of a master equation \cite{AvLikPaper, AverinLikharevBookChapter, Beenakker, Kulik, Ben-JacobWilkins, AvLikKor}} involving electrons of all energies. In the non-thermalized regime (\ref{g<G}), the electrons with different energies are not mixed. Hence, the probabilities and distribution functions can be found from the energy-conserving quantum kinetic equation (QKE), which is formulated using the Keldysh formalism (see, e.g.,  \cite{MeirWingreenJauho,RammerSmith,HaugJauho}) in terms of the ``greater'', $g^>(t)$, and {``lesser'', $g^<(t)$, Green's function} of the dot.
In the regime (\ref{scales}), where the mean level spacing is much larger than the level broadening due to {tunneling}, they are split into a sum over the energy levels, with Green's function for the $n^\mathrm{th}$ level given by $g^>_n(t) = -i \langle d_n(t)d_n^\dagger(0) \rangle$ and $g^<_n(t) = i \langle d_n^\dagger(0)  d_n(t) \rangle$, where $d_n(t) = \mathrm{e}^{iHt}d_n \mathrm{e}^{-iHt}$. Then, to linear order in tunneling, the QKE is reduced to \cite{MeirWingreenJauho,RammerSmith,HaugJauho},
\begin{equation}\label{QKE Simple}
    g_n^>(\varepsilon)\Sigma^<(\varepsilon) = g_n^<(\varepsilon) \Sigma^>(\varepsilon).
\end{equation}
Here, {the conservation} of particle number for an isolated dot allows one to represent the single-level Green's functions as (see \emph{Supplemental Material)}
\begin{align} \label{g<g>}
    g^>_n(\varepsilon) &= -2\pi i \sum_N \delta \left(\varepsilon - \varepsilon_n - \Omega_N \right) p_N(1-F_N(\varepsilon_n)), \notag\\
    g^<_n(\varepsilon) &= 2\pi i \sum_N \delta \left(\varepsilon - \varepsilon_n - \Omega_{N-1} \right) p_N F_N(\varepsilon_n),
\end{align}
with the normalization $\sum_N p_N = 1$.  The self-energy functions of the leads in Eq.~\eqref{QKE Simple} are assumed to be $n$-independent  and are given by
    \begin{align}\label{Sigma}
\Sigma^>(\varepsilon) &=  i \sum_{\alpha ={\mathrm{L,R}}}\Gamma_\alpha (f_\alpha(\varepsilon){-}1),&\Sigma^<(\varepsilon) &= i \sum_{\alpha ={\mathrm{L,R}}} \Gamma_\alpha f_\alpha(\varepsilon).
\end{align}
 Substituting Eqs.~(\ref{g<g>})and  (\ref{Sigma}) into Eq.~(\ref{QKE Simple}) leads to the QKE reflecting the detailed balance equations,  {coinciding with those derived} in \cite{Beenakker} for $\Delta \gg T$,
 \begin{align}\label{QKE}
    p_{N+1}F_{N+1}(\varepsilon_n)&\left(1-\widetilde{f}(\varepsilon_n + \Omega_N) \right)\notag
	\\&= p_N \left(1-F_N(\varepsilon_n)\right)\widetilde{f}(\varepsilon_n + \Omega_N),\\
\widetilde{f}(\varepsilon)& = (\Gamma _{\mathrm{L}}/\Gamma)f_{\mathrm{L}}(\varepsilon) + (\Gamma _{\mathrm{R}}/\Gamma)f_{\mathrm{R}}(\varepsilon) .\notag
\end{align}
 It is this equation along with the normalization conditions, $\sum_N p_N =1$ and $\sum_n F_N(\varepsilon_n) = N$, that can be used to obtain the probabilities and distribution functions required in Eq.~(\ref{Current}) to calculate the current. The results for full thermalization are recovered by summing Eq.~(\ref{QKE}) over $n$ using the fact that in this case we can substitute the equilibrium distribution function,
  $F_N(\varepsilon_n) = f(\varepsilon_n {-} \varepsilon_\mathrm{F})$.

The absence of thermalization, however, drastically changes the distribution function. In this case, QKE (\ref{QKE}) has an exact solution providing there are only two relevant states ($N$ and $N+1$) for a given voltage (see \emph{Supplemental Material}).    In the case of approximately equal coupling, this condition can be satisfied only for a finite bias window, i.e.\ within the first step of the Coulomb staircase. For higher bias, one needs to account for more states with different numbers of particles that are not being exponentially suppressed (in contrast to the asymmetric case when $\Gamma_{\mathrm{L}}/\Gamma_{\mathrm{R}}\gg1$ \cite{AsymCBPaper}).

Assuming that the chemical potential in the dot is of order of the unbiased chemical potential in the (right) lead, we show that the current and, hence, the differential conductance has an additional peak in the window   $0 \leq eV \lesssim \Omega_{N+1}$ as compared to the thermalized case \cite{AverinLikharevBookChapter, Kulik, Ben-JacobWilkins, AvLikKor}. In this window, where only two levels are relevant,  the kinetic equation (\ref{QKE}) has the solution $F_N(\varepsilon_n) \approx F_{N+1}(\varepsilon_n) \approx F(\varepsilon_n)$ in the limit $N \gg 1$, leading to
\begin{equation} \label{QKE F Soln}
	F(\varepsilon_n)  = \frac{\widetilde{f}(\varepsilon_n + \Omega_N)}{[1-\widetilde{f}(\varepsilon_n + \Omega_N)]A_N+\widetilde{f}(\varepsilon_n + \Omega_N)},
\end{equation}
where $A_N = p_{N+1}/p_N$. This ratio of probabilities is found from normalization, $N = \sum_n F(\varepsilon_n) =  (1/\Delta)\int_0^{\infty} F(\varepsilon)\mathrm{d}\varepsilon$, while $p_N + p_{N+1}=1$ as shown in \emph{Supplemental Material}. As seen in Fig.~(\ref{Fig:probs}), depicted for the middle of the Coulomb blockade valley where $\Omega_N=E_{\mathrm{c}}/2$, both $p_N$ and $p_{N+1}$ are practically indistinguishable from the thermalized case. It remains the case as long as $\Omega_N$, Eq.~\eqref{OmegaN}, remains far from the peaks of the Coulomb blockade. Note that this and all subsequent results depend only on ratios of energetic parameters and are fully applicable in experimental regimes where $T\sim 10-100{\mathrm{mK}}$ and $E_c\sim 1{\mathrm{meV}}$.

\begin{figure}[t]
	\includegraphics[width = \columnwidth]{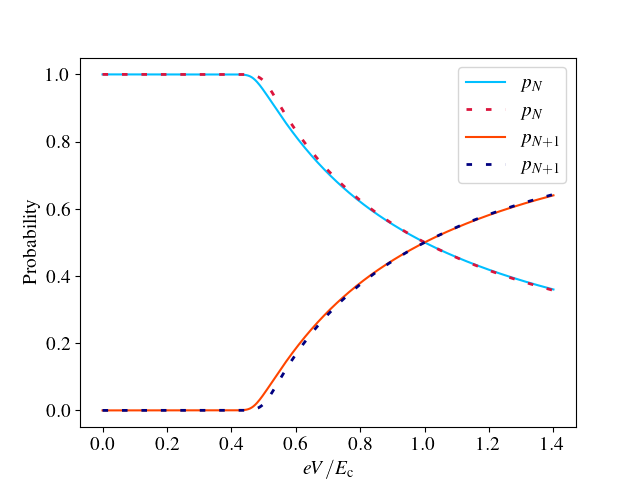}
	\caption{\label{Fig:probs}The occupation probabilities $p_N$ (the upper line) and $p_{N{+}1}$ (the lower line) as functions of bias voltage, $V$, for $\Gamma_\mathrm{L} {=} \Gamma_\mathrm{R}$, and $ N {=} N_{\mathrm{g}}$. Here they depend only on the ratio $E_{\mathrm{c}}/T$ in the temperature region $10-100$mK, albeit this dependence is rather weak ($  E_{\mathrm{c}}/T=100$ was used for this figure).The solid lines represent the results for the non-thermalized case, the dashed lines  for the full-thermalization  case \cite{AverinLikharevBookChapter, Kulik, Ben-JacobWilkins, AvLikKor}.  In this temperature range they are  practically indistinguishable.}
\end{figure}

\begin{figure}[t]
	\includegraphics[width=\columnwidth]{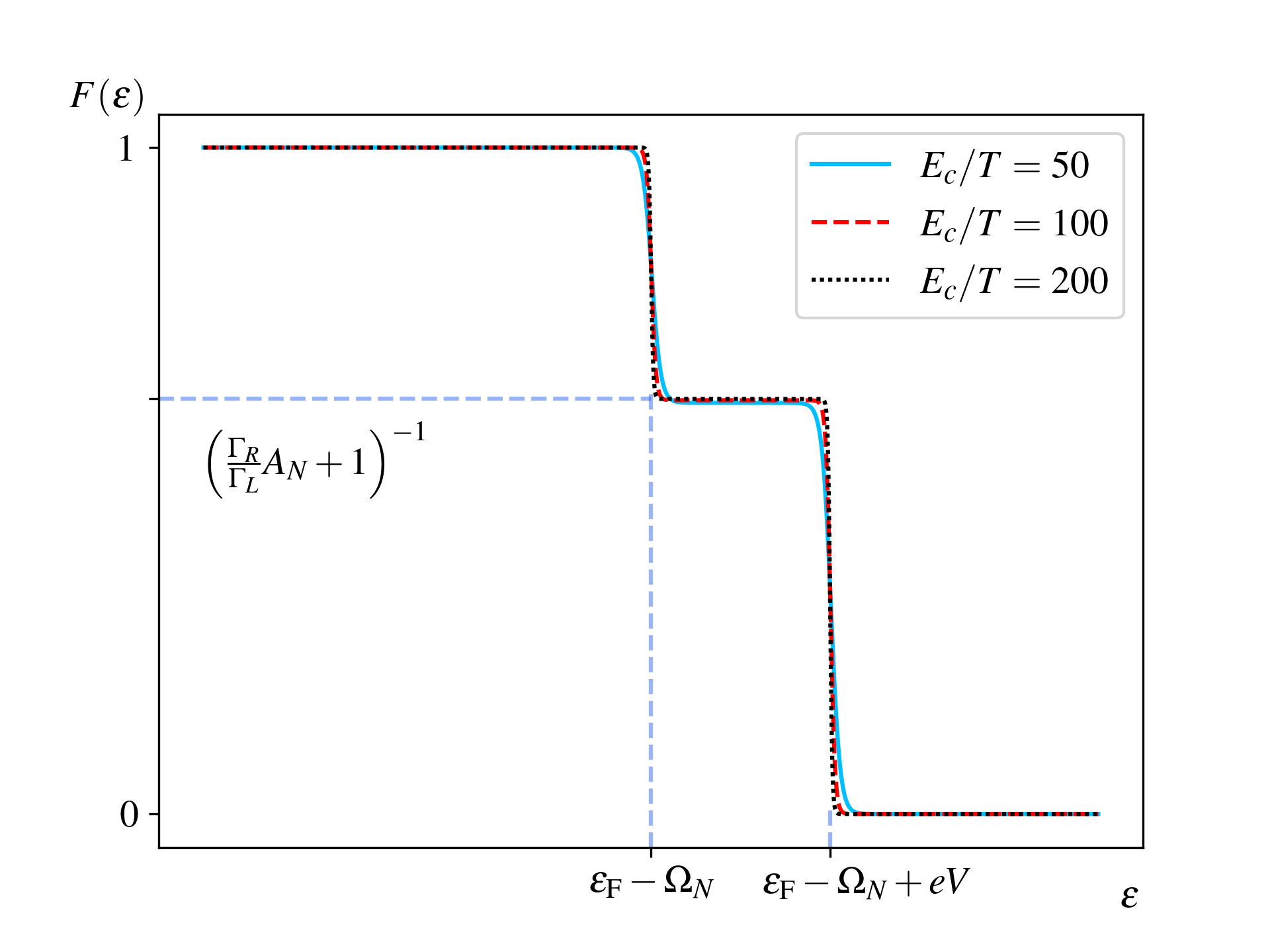}
	\caption{\label{Fig:FN}The electron distribution function in the dot   for $\Gamma_{\mathrm{L}}=\Gamma_{\mathrm{R}}$, $N=N_{\mathrm{g}}$, and $eV =0.8E_{\mathrm{c}}$ where numerically we find $A_N\approx0.6$. The double-step structure is robust as long as $eV>\Omega_N$ -- in the opposite case $A_N\equiv p_{N{+1}}/p_N\to0$, as seen from Fig.~(\ref{Fig:probs}), and the middle step disappears. $F\qty(\varepsilon )$ has only a weak dependence on $E_{\mathrm{c}}/T$ so that the three curves above practically merge.}
\end{figure}

  On the contrary, the distribution function, found by substituting the  ratio $A_N \equiv p_{N+1}/p_N$ into Eq.~\eqref{QKE F Soln},   acquires an additional step
\begin{equation}\label{FN regions}
	F(\varepsilon_n) \approx \begin{cases}
		1, &  \varepsilon_n {<} \mu_{\mathrm{R}}{-}\Omega_N\\
		\left(1{+}\frac{\Gamma_\mathrm{R}}{\Gamma_\mathrm{L}}A_N \right)^{\!{-}1}\!\!\!,  &\mu_{\mathrm{R}}{-}\Omega_N   {<}\varepsilon_n {<} \mu_{\mathrm{L}}{-}\Omega_N \\
		0. & \mu_{\mathrm{L}} {-}\Omega_N {<}\varepsilon_n
	\end{cases}
\end{equation}
as depicted for the middle of the valley in Fig.~\ref{Fig:FN}. Such a double-step is similar to that observed in short quasi-one-dimensional wires \cite{Birge}. However, in the wire the double-step was simply a linear combination of the two Fermi-functions of the leads, while in the present case it is substantially affected by the Coulomb interaction. Still, in both cases the double-step reflects the lack of thermalization between electrons coming from the left and right leads. In the steady-state limit, electrons from both leads enter the dot at two different chemical potentials and thermalize with the opposite lead only after exiting the dot. Note that the double-step is effectively washed out in the one-lead limit of the Coulomb blockade when $\Gamma_{\mathrm{R}}/\Gamma_{\mathrm{L}}\ll1$.

\begin{figure}[t]
	\includegraphics[width=\columnwidth]{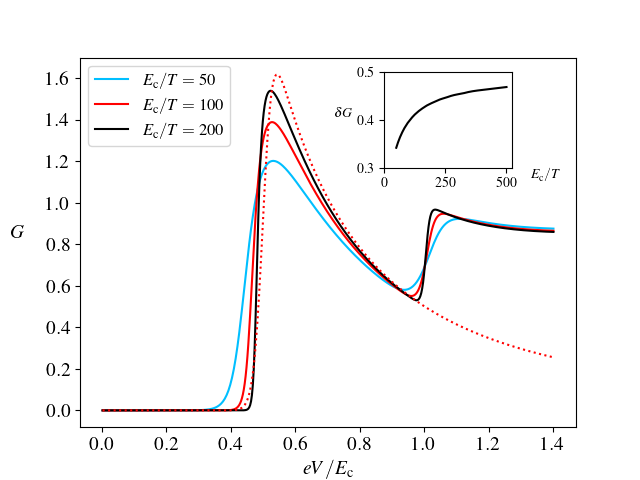}
	\caption{\label{Fig:Conductance}The differential conductance, $G(V)$,  in units of $\frac{e^2}{\Delta}\frac{\Gamma_\mathrm{L}\Gamma_\mathrm{R}}{\Gamma}\approx \frac{e^2\Gamma}{4\Delta}$ when    $\Gamma_\mathrm{L} \approx \Gamma_\mathrm{R}$. For $\Gamma_{\mathrm{L}}\sim\Gamma_{\mathrm{R}}$, an additional jump in $G$ is always at $eV=E_{\mathrm{c}}$. Such a jump is absent in the thermalized case {\cite{AverinLikharevBookChapter, Kulik, Ben-JacobWilkins, AvLikKor}}, depicted here by the dotted line for $E_{\mathrm{c}}/T=100$. The dependence of its height, $\delta G$,   on the ratio $E_{\mathrm{c}}/T$, is shown in the inset.}
\end{figure}

The double-step distribution in the dot drastically changes  the differential conductance, $G= {\mathrm{d}I}/{\mathrm{d}V}$, in comparison with the thermalized case {\cite{AverinLikharevBookChapter, Kulik, Ben-JacobWilkins, AvLikKor}}.
Substituting $p_N$ and $F(\varepsilon_n)$ into Eq.~(\ref{Current}) with $F_N(\varepsilon_n) \approx F(\varepsilon_n)$, we find $G$ as shown in Fig.~\ref{Fig:Conductance}. For small voltages, $eV < E_\mathrm{c}$, the absence of thermalization has little impact in the low-$T$ limit. However, at $eV = E_\mathrm{c}$, there appears a secondary jump in the non-thermalized case. It is robust as long as the tunneling is symmetric, $\Gamma_\mathrm{L} \approx \Gamma_\mathrm{R}$, when there are three distinct regions for the distribution, Eq.~(\ref{FN regions}). Rewriting Eq.~(\ref{Current}) for the current in the low-$T$ limit and for  $eV \lesssim \Omega_{N+1}$  will make this clearer:
\begin{multline}\label{Current Modified}
		I =  \frac{e}\Delta\frac{\Gamma_\mathrm{L} \Gamma_\mathrm{R}} {\Gamma }  \Big( p_N \int_{\mu-\Omega_{N-1}}^{\mu-\Omega_{N-1}+eV} F(\varepsilon) \mathrm{d}\varepsilon \\ + \int_{\mu -\Omega_N}^{\mu -\Omega_N+eV} \big[p_N (1-F(\varepsilon)) + p_{N+1}F(\varepsilon)\big] \mathrm{d}\varepsilon \\+ p_{N+1} \int_{\mu -\Omega_{N+1}}^{\mu -\Omega_{N+1}+eV} (1-F(\varepsilon)) \mathrm{d}\varepsilon \Big).
\end{multline}
The second integration over the middle step starts to contribute at $eV\geqslant E_\mathrm{c}/2$ when $p_N$ and $p_{N{+1}}$ start to change, see Fig.~\ref{Fig:probs}, signaling that the oncoming particle is sufficiently energetic to overcome the charging energy. This results in the usual blockade jump  which is the same for both the thermalized and non-thermalized cases. As long as $eV<E_\mathrm{c}$, the first and third integrals in Eq.~\eqref{Current Modified} are negligible as the each integration is over a region where the integrands are exponentially small at $T\ll E_\mathrm{c}$. For $eV\geqslant E_\mathrm{c}$, this is no longer the case and the appropriate non-zero contribution results in a sudden change in the current revealed as a jump in the differential conductance at $eV = E_\mathrm{c}$.

The position of this jump is insensitive to gate voltage as it only depends on the difference  $\Omega_{N+1}-\Omega_{N}=E_\mathrm{c}$.  In the region around the jump, the ratio of probabilities is given for $T\ll E_{\mathrm{c}}$ (see \emph{Supplemental Material}) by
\begin{equation}\label{Jump AN}
	A_N \equiv \frac{p_{N+1}}{p_N}\approx \frac{\Gamma_\mathrm{L}}{\Gamma_\mathrm{R}}\left(\frac{eV-\Omega_N}{\Omega_N} \right).
\end{equation}
Then, calculating the current from Eq.~(\ref{Current Modified}) on both sides of the jump we find  that the jump in the differential conductance, neglecting corrections in $T/E_{\mathrm{c}}$, has the height
\begin{equation}\label{G jump}
	\delta G = \frac{e^2}{2\Delta}\frac{\Gamma_\mathrm{L}\Gamma_\mathrm{R}}{\Gamma},
\end{equation}
in the middle of the Coulomb blockade valley, $\Omega_N=\frac12E_{\mathrm{c}}$. (The general expression for $\delta G$ is given in \emph{Supplemental Material}). %
This jump is rather robust: it occurs at $eV=E_{\mathrm{c}}$ independently of $\Omega_N$ and has only a weak temperature dependence. As the temperature is increased, while still obeying inequalities (\ref{scales}), the jump is only slightly smeared across a wider range of voltages as shown in the inset in Fig.~(\ref{Fig:Conductance}). This jump should be experimentally observable and give a clear indication of the absence of thermalization within a quantum dot.

In conclusion we note that the existence of additional fine structure of the Coulomb blockade peaks has been established numerically and experimentally \cite{AltshulerTinkham} for small dots, where $\Delta\gg T$. Here we  have considered the opposite case of large quantum dots, \eqref{scales}, where we have shown that  the absence of thermalization    manifests itself as an additional jump in the differential conductance at $eV=E_\mathrm{c}$, which follows the usual jump at  $eV=\Omega_N$. This is a direct consequence of the lack of equilibration between electrons coming from the left and right leads so that the distribution function on the dot has a double-step form. We anticipate this jump to be experimentally accessible at the appropriate voltages  and therefore could be used as a method of identifying the absence of thermalization in the dot.

\begin{acknowledgements}
	We gratefully acknowledge support from EPSRC  under the grant EP/R029075/1 (IVL) and from the Leverhulme Trust under the grant  RPG-2019-317 (IVY).
\end{acknowledgements}

\bibliography{bibliography.bib}

\begin{thebibliography}{28}%
\makeatletter
\providecommand \@ifxundefined [1]{%
 \@ifx{#1\undefined}
}%
\providecommand \@ifnum [1]{%
 \ifnum #1\expandafter \@firstoftwo
 \else \expandafter \@secondoftwo
 \fi
}%
\providecommand \@ifx [1]{%
 \ifx #1\expandafter \@firstoftwo
 \else \expandafter \@secondoftwo
 \fi
}%
\providecommand \natexlab [1]{#1}%
\providecommand \enquote  [1]{``#1''}%
\providecommand \bibnamefont  [1]{#1}%
\providecommand \bibfnamefont [1]{#1}%
\providecommand \citenamefont [1]{#1}%
\providecommand \href@noop [0]{\@secondoftwo}%
\providecommand \href [0]{\begingroup \@sanitize@url \@href}%
\providecommand \@href[1]{\@@startlink{#1}\@@href}%
\providecommand \@@href[1]{\endgroup#1\@@endlink}%
\providecommand \@sanitize@url [0]{\catcode `\\12\catcode `\$12\catcode `\&12\catcode `\#12\catcode `\^12\catcode `\_12\catcode `\%12\relax}%
\providecommand \@@startlink[1]{}%
\providecommand \@@endlink[0]{}%
\providecommand \url  [0]{\begingroup\@sanitize@url \@url }%
\providecommand \@url [1]{\endgroup\@href {#1}{\urlprefix }}%
\providecommand \urlprefix  [0]{URL }%
\providecommand \Eprint [0]{\href }%
\providecommand \doibase [0]{https://doi.org/}%
\providecommand \selectlanguage [0]{\@gobble}%
\providecommand \bibinfo  [0]{\@secondoftwo}%
\providecommand \bibfield  [0]{\@secondoftwo}%
\providecommand \translation [1]{[#1]}%
\providecommand \BibitemOpen [0]{}%
\providecommand \bibitemStop [0]{}%
\providecommand \bibitemNoStop [0]{.\EOS\space}%
\providecommand \EOS [0]{\spacefactor3000\relax}%
\providecommand \BibitemShut  [1]{\csname bibitem#1\endcsname}%
\let\auto@bib@innerbib\@empty
\bibitem [{\citenamefont {Basko}\ \emph {et~al.}(2006)\citenamefont {Basko}, \citenamefont {Aleiner},\ and\ \citenamefont {Altshuler}}]{BAA2006}%
  \BibitemOpen
  \bibfield  {author} {\bibinfo {author} {\bibfnamefont {D.}~\bibnamefont {Basko}}, \bibinfo {author} {\bibfnamefont {I.}~\bibnamefont {Aleiner}},\ and\ \bibinfo {author} {\bibfnamefont {B.}~\bibnamefont {Altshuler}},\ }\bibfield  {title} {\bibinfo {title} {Metal--insulator transition in a weakly interacting many-electron system with localized single-particle states},\ }\href {https://doi.org/https://doi.org/10.1016/j.aop.2005.11.014} {\bibfield  {journal} {\bibinfo  {journal} {Annals of Physics}\ }\textbf {\bibinfo {volume} {321}},\ \bibinfo {pages} {1126} (\bibinfo {year} {2006})}\BibitemShut {NoStop}%
\bibitem [{\citenamefont {Gornyi}\ \emph {et~al.}(2005)\citenamefont {Gornyi}, \citenamefont {Mirlin},\ and\ \citenamefont {Polyakov}}]{MBL2}%
  \BibitemOpen
  \bibfield  {author} {\bibinfo {author} {\bibfnamefont {I.~V.}\ \bibnamefont {Gornyi}}, \bibinfo {author} {\bibfnamefont {A.~D.}\ \bibnamefont {Mirlin}},\ and\ \bibinfo {author} {\bibfnamefont {D.~G.}\ \bibnamefont {Polyakov}},\ }\bibfield  {title} {\bibinfo {title} {Interacting electrons in disordered wires: Anderson localization and low-$t$ transport},\ }\href {https://doi.org/10.1103/PhysRevLett.95.206603} {\bibfield  {journal} {\bibinfo  {journal} {Phys. Rev. Lett.}\ }\textbf {\bibinfo {volume} {95}},\ \bibinfo {pages} {206603} (\bibinfo {year} {2005})}\BibitemShut {NoStop}%
\bibitem [{\citenamefont {Abanin}\ and\ \citenamefont {Papi{\'c}}(2017)}]{MBLReviewPapic}%
  \BibitemOpen
  \bibfield  {author} {\bibinfo {author} {\bibfnamefont {D.~A.}\ \bibnamefont {Abanin}}\ and\ \bibinfo {author} {\bibfnamefont {Z.}~\bibnamefont {Papi{\'c}}},\ }\bibfield  {title} {\bibinfo {title} {Recent progress in many-body localization},\ }\href {https://doi.org/https://doi.org/10.1002/andp.201700169} {\bibfield  {journal} {\bibinfo  {journal} {Annalen der Physik}\ }\textbf {\bibinfo {volume} {529}},\ \bibinfo {pages} {1700169} (\bibinfo {year} {2017})}\BibitemShut {NoStop}%
\bibitem [{\citenamefont {Schreiber}\ \emph {et~al.}(2015)\citenamefont {Schreiber}, \citenamefont {Hodgman}, \citenamefont {Bordia}, \citenamefont {L{\"u}schen}, \citenamefont {Fischer}, \citenamefont {Vosk}, \citenamefont {Altman}, \citenamefont {Schneider},\ and\ \citenamefont {Bloch}}]{MBLExpt}%
  \BibitemOpen
  \bibfield  {author} {\bibinfo {author} {\bibfnamefont {M.}~\bibnamefont {Schreiber}}, \bibinfo {author} {\bibfnamefont {S.~S.}\ \bibnamefont {Hodgman}}, \bibinfo {author} {\bibfnamefont {P.}~\bibnamefont {Bordia}}, \bibinfo {author} {\bibfnamefont {H.~P.}\ \bibnamefont {L{\"u}schen}}, \bibinfo {author} {\bibfnamefont {M.~H.}\ \bibnamefont {Fischer}}, \bibinfo {author} {\bibfnamefont {R.}~\bibnamefont {Vosk}}, \bibinfo {author} {\bibfnamefont {E.}~\bibnamefont {Altman}}, \bibinfo {author} {\bibfnamefont {U.}~\bibnamefont {Schneider}},\ and\ \bibinfo {author} {\bibfnamefont {I.}~\bibnamefont {Bloch}},\ }\bibfield  {title} {\bibinfo {title} {Observation of many-body localization of interacting fermions in a quasirandom optical lattice},\ }\href {https://doi.org/10.1126/science.aaa7432} {\bibfield  {journal} {\bibinfo  {journal} {Science}\ }\textbf {\bibinfo {volume} {349}},\ \bibinfo {pages} {842} (\bibinfo {year} {2015})}\BibitemShut {NoStop}%
\bibitem [{\citenamefont {Nandkishore}\ and\ \citenamefont {Huse}(2015)}]{NandkishoreHuse}%
  \BibitemOpen
  \bibfield  {author} {\bibinfo {author} {\bibfnamefont {R.}~\bibnamefont {Nandkishore}}\ and\ \bibinfo {author} {\bibfnamefont {D.~A.}\ \bibnamefont {Huse}},\ }\bibfield  {title} {\bibinfo {title} {Many-body localization and thermalization in quantum statistical mechanics},\ }\href {https://doi.org/10.1146/annurev-conmatphys-031214-014726} {\bibfield  {journal} {\bibinfo  {journal} {Annual Review of Condensed Matter Physics}\ }\textbf {\bibinfo {volume} {6}},\ \bibinfo {pages} {15} (\bibinfo {year} {2015})}\BibitemShut {NoStop}%
\bibitem [{\citenamefont {Serbyn}\ \emph {et~al.}(2021)\citenamefont {Serbyn}, \citenamefont {Abanin},\ and\ \citenamefont {Papi{\'c}}}]{ScarsReview}%
  \BibitemOpen
  \bibfield  {author} {\bibinfo {author} {\bibfnamefont {M.}~\bibnamefont {Serbyn}}, \bibinfo {author} {\bibfnamefont {D.~A.}\ \bibnamefont {Abanin}},\ and\ \bibinfo {author} {\bibfnamefont {Z.}~\bibnamefont {Papi{\'c}}},\ }\bibfield  {title} {\bibinfo {title} {Quantum many-body scars and weak breaking of ergodicity},\ }\href {https://doi.org/10.1038/s41567-021-01230-2} {\bibfield  {journal} {\bibinfo  {journal} {Nature Physics}\ }\textbf {\bibinfo {volume} {17}},\ \bibinfo {pages} {675} (\bibinfo {year} {2021})}\BibitemShut {NoStop}%
\bibitem [{\citenamefont {Altshuler}\ \emph {et~al.}(1997)\citenamefont {Altshuler}, \citenamefont {Gefen}, \citenamefont {Kamenev},\ and\ \citenamefont {Levitov}}]{AKGL}%
  \BibitemOpen
  \bibfield  {author} {\bibinfo {author} {\bibfnamefont {B.~L.}\ \bibnamefont {Altshuler}}, \bibinfo {author} {\bibfnamefont {Y.}~\bibnamefont {Gefen}}, \bibinfo {author} {\bibfnamefont {A.}~\bibnamefont {Kamenev}},\ and\ \bibinfo {author} {\bibfnamefont {L.~S.}\ \bibnamefont {Levitov}},\ }\bibfield  {title} {\bibinfo {title} {Quasiparticle lifetime in a finite system: a nonperturbative approach},\ }\href {https://doi.org/10.1103/PhysRevLett.78.2803} {\bibfield  {journal} {\bibinfo  {journal} {Phys. Rev. Lett.}\ }\textbf {\bibinfo {volume} {78}},\ \bibinfo {pages} {2803} (\bibinfo {year} {1997})}\BibitemShut {NoStop}%
\bibitem [{\citenamefont {Sivan}\ \emph {et~al.}(1994)\citenamefont {Sivan}, \citenamefont {Imry},\ and\ \citenamefont {Aronov}}]{SivanImry1994}%
  \BibitemOpen
  \bibfield  {author} {\bibinfo {author} {\bibfnamefont {U.}~\bibnamefont {Sivan}}, \bibinfo {author} {\bibfnamefont {Y.}~\bibnamefont {Imry}},\ and\ \bibinfo {author} {\bibfnamefont {A.~G.}\ \bibnamefont {Aronov}},\ }\bibfield  {title} {\bibinfo {title} {Quasi-particle lifetime in a quantum dot},\ }\href {https://doi.org/10.1209/0295-5075/28/2/007} {\bibfield  {journal} {\bibinfo  {journal} {Europhys. Lett.}\ }\textbf {\bibinfo {volume} {28}},\ \bibinfo {pages} {115} (\bibinfo {year} {1994})}\BibitemShut {NoStop}%
\bibitem [{\citenamefont {Blanter}(1996)}]{BlanterRates}%
  \BibitemOpen
  \bibfield  {author} {\bibinfo {author} {\bibfnamefont {Y.~M.}\ \bibnamefont {Blanter}},\ }\bibfield  {title} {\bibinfo {title} {Electron-electron scattering rate in disordered mesoscopic systems},\ }\href {https://doi.org/10.1103/PhysRevB.54.12807} {\bibfield  {journal} {\bibinfo  {journal} {Phys. Rev. B}\ }\textbf {\bibinfo {volume} {54}},\ \bibinfo {pages} {12807} (\bibinfo {year} {1996})}\BibitemShut {NoStop}%
\bibitem [{\citenamefont {Kouwenhoven}\ \emph {et~al.}(1997)\citenamefont {Kouwenhoven}, \citenamefont {Marcus}, \citenamefont {McEuen}, \citenamefont {Tarucha}, \citenamefont {Westervelt},\ and\ \citenamefont {Wingreen}}]{Kouwenhoven1997}%
  \BibitemOpen
  \bibfield  {author} {\bibinfo {author} {\bibfnamefont {L.~P.}\ \bibnamefont {Kouwenhoven}}, \bibinfo {author} {\bibfnamefont {C.~M.}\ \bibnamefont {Marcus}}, \bibinfo {author} {\bibfnamefont {P.~L.}\ \bibnamefont {McEuen}}, \bibinfo {author} {\bibfnamefont {S.}~\bibnamefont {Tarucha}}, \bibinfo {author} {\bibfnamefont {R.~M.}\ \bibnamefont {Westervelt}},\ and\ \bibinfo {author} {\bibfnamefont {N.~S.}\ \bibnamefont {Wingreen}},\ }\bibfield  {title} {\bibinfo {title} {Electron transport in quantum dots},\ }in\ \href@noop {} {\emph {\bibinfo {booktitle} {Mesoscopic Electron Transport}}},\ \bibinfo {editor} {edited by\ \bibinfo {editor} {\bibfnamefont {L.~L.}\ \bibnamefont {Sohn}}, \bibinfo {editor} {\bibfnamefont {L.~P.}\ \bibnamefont {Kouwenhoven}},\ and\ \bibinfo {editor} {\bibfnamefont {G.}~\bibnamefont {Sch{\"o}n}}}\ (\bibinfo  {publisher} {Springer Netherlands},\ \bibinfo {address} {Dordrecht},\ \bibinfo {year} {1997})\ pp.\ \bibinfo {pages} {105--214}\BibitemShut {NoStop}%
\bibitem [{\citenamefont {Aleiner}\ \emph {et~al.}(2002)\citenamefont {Aleiner}, \citenamefont {Brouwer},\ and\ \citenamefont {Glazman}}]{AleinerReview}%
  \BibitemOpen
  \bibfield  {author} {\bibinfo {author} {\bibfnamefont {I.~L.}\ \bibnamefont {Aleiner}}, \bibinfo {author} {\bibfnamefont {P.~W.}\ \bibnamefont {Brouwer}},\ and\ \bibinfo {author} {\bibfnamefont {L.~I.}\ \bibnamefont {Glazman}},\ }\bibfield  {title} {\bibinfo {title} {Quantum effects in coulomb blockade},\ }\href {https://doi.org/https://doi.org/10.1016/S0370-1573(01)00063-1} {\bibfield  {journal} {\bibinfo  {journal} {Phys. Rep.}\ }\textbf {\bibinfo {volume} {358}},\ \bibinfo {pages} {309} (\bibinfo {year} {2002})}\BibitemShut {NoStop}%
\bibitem [{\citenamefont {Alhassid}(2000)}]{AlhassidReview}%
  \BibitemOpen
  \bibfield  {author} {\bibinfo {author} {\bibfnamefont {Y.}~\bibnamefont {Alhassid}},\ }\bibfield  {title} {\bibinfo {title} {The statistical theory of quantum dots},\ }\href {https://doi.org/10.1103/RevModPhys.72.895} {\bibfield  {journal} {\bibinfo  {journal} {Rev. Mod. Phys.}\ }\textbf {\bibinfo {volume} {72}},\ \bibinfo {pages} {895} (\bibinfo {year} {2000})}\BibitemShut {NoStop}%
\bibitem [{\citenamefont {Averin}\ and\ \citenamefont {Likharev}(1986)}]{AvLikPaper}%
  \BibitemOpen
  \bibfield  {author} {\bibinfo {author} {\bibfnamefont {D.~V.}\ \bibnamefont {Averin}}\ and\ \bibinfo {author} {\bibfnamefont {K.~K.}\ \bibnamefont {Likharev}},\ }\bibfield  {title} {\bibinfo {title} {Coulomb blockade of single-electron tunneling, and coherent oscillations in small tunnel junctions},\ }\href {https://doi.org/10.1007/BF00683469} {\bibfield  {journal} {\bibinfo  {journal} {Journal of Low Temperature Physics}\ }\textbf {\bibinfo {volume} {62}},\ \bibinfo {pages} {345} (\bibinfo {year} {1986})}\BibitemShut {NoStop}%
\bibitem [{\citenamefont {Beenakker}(1991)}]{Beenakker}%
  \BibitemOpen
  \bibfield  {author} {\bibinfo {author} {\bibfnamefont {C.~W.~J.}\ \bibnamefont {Beenakker}},\ }\bibfield  {title} {\bibinfo {title} {Theory of coulomb-blockade oscillations in the conductance of a quantum dot},\ }\href {https://doi.org/10.1103/PhysRevB.44.1646} {\bibfield  {journal} {\bibinfo  {journal} {Phys. Rev. B}\ }\textbf {\bibinfo {volume} {44}},\ \bibinfo {pages} {1646} (\bibinfo {year} {1991})}\BibitemShut {NoStop}%
\bibitem [{\citenamefont {Sedlmayr}\ \emph {et~al.}(2006)\citenamefont {Sedlmayr}, \citenamefont {Yurkevich},\ and\ \citenamefont {Lerner}}]{TDoS}%
  \BibitemOpen
  \bibfield  {author} {\bibinfo {author} {\bibfnamefont {N.}~\bibnamefont {Sedlmayr}}, \bibinfo {author} {\bibfnamefont {I.~V.}\ \bibnamefont {Yurkevich}},\ and\ \bibinfo {author} {\bibfnamefont {I.~V.}\ \bibnamefont {Lerner}},\ }\bibfield  {title} {\bibinfo {title} {Tunnelling density of states at coulomb-blockade peaks},\ }\href {https://doi.org/10.1209/epl/i2006-10236-0} {\bibfield  {journal} {\bibinfo  {journal} {Europhys. Lett.}\ }\textbf {\bibinfo {volume} {76}},\ \bibinfo {pages} {109} (\bibinfo {year} {2006})}\BibitemShut {NoStop}%
\bibitem [{\citenamefont {Kulik}\ and\ \citenamefont {Shekhter}(1975)}]{Kulik}%
  \BibitemOpen
  \bibfield  {author} {\bibinfo {author} {\bibfnamefont {I.~O.}\ \bibnamefont {Kulik}}\ and\ \bibinfo {author} {\bibfnamefont {R.~I.}\ \bibnamefont {Shekhter}},\ }\bibfield  {title} {\bibinfo {title} {Kinetic phenomena and charge discreteness effects in granulated media},\ }\href@noop {} {\bibfield  {journal} {\bibinfo  {journal} {Zh. Eksp. Teor. Fiz.}\ }\textbf {\bibinfo {volume} {68}},\ \bibinfo {pages} {623} (\bibinfo {year} {1975})}\BibitemShut {NoStop}%
\bibitem [{\citenamefont {Averin}\ and\ \citenamefont {Likharev}(1991)}]{AverinLikharevBookChapter}%
  \BibitemOpen
  \bibfield  {author} {\bibinfo {author} {\bibfnamefont {D.}~\bibnamefont {Averin}}\ and\ \bibinfo {author} {\bibfnamefont {K.}~\bibnamefont {Likharev}},\ }\bibfield  {title} {\bibinfo {title} {A correlated transfer of single electrons and {C}ooper pairs in systems of small tunnel junctions},\ }in\ \href {https://doi.org/https://doi.org/10.1016/B978-0-444-88454-1.50012-7} {\emph {\bibinfo {booktitle} {Mesoscopic Phenomena in Solids}}},\ \bibinfo {series} {Modern Problems in Condensed Matter Sciences}, Vol.~\bibinfo {volume} {30},\ \bibinfo {editor} {edited by\ \bibinfo {editor} {\bibfnamefont {B.}~\bibnamefont {Altshuler}}, \bibinfo {editor} {\bibfnamefont {P.}~\bibnamefont {Lee}},\ and\ \bibinfo {editor} {\bibfnamefont {R.}~\bibnamefont {Webb}}}\ (\bibinfo  {publisher} {Elsevier},\ \bibinfo {address} {Amsterdam},\ \bibinfo {year} {1991})\ pp.\ \bibinfo {pages} {173--271}\BibitemShut {NoStop}%
\bibitem [{\citenamefont {Amman}\ \emph {et~al.}(1991)\citenamefont {Amman}, \citenamefont {Wilkins}, \citenamefont {Ben-Jacob}, \citenamefont {Maker},\ and\ \citenamefont {Jaklevic}}]{Ben-JacobWilkins}%
  \BibitemOpen
  \bibfield  {author} {\bibinfo {author} {\bibfnamefont {M.}~\bibnamefont {Amman}}, \bibinfo {author} {\bibfnamefont {R.}~\bibnamefont {Wilkins}}, \bibinfo {author} {\bibfnamefont {E.}~\bibnamefont {Ben-Jacob}}, \bibinfo {author} {\bibfnamefont {P.~D.}\ \bibnamefont {Maker}},\ and\ \bibinfo {author} {\bibfnamefont {R.~C.}\ \bibnamefont {Jaklevic}},\ }\bibfield  {title} {\bibinfo {title} {Analytic solution for the current-voltage characteristic of two mesoscopic tunnel junctions coupled in series},\ }\href {https://doi.org/10.1103/PhysRevB.43.1146} {\bibfield  {journal} {\bibinfo  {journal} {Phys. Rev. B}\ }\textbf {\bibinfo {volume} {43}},\ \bibinfo {pages} {1146} (\bibinfo {year} {1991})}\BibitemShut {NoStop}%
\bibitem [{\citenamefont {Averin}\ \emph {et~al.}(1991)\citenamefont {Averin}, \citenamefont {Korotkov},\ and\ \citenamefont {Likharev}}]{AvLikKor}%
  \BibitemOpen
  \bibfield  {author} {\bibinfo {author} {\bibfnamefont {D.~V.}\ \bibnamefont {Averin}}, \bibinfo {author} {\bibfnamefont {A.~N.}\ \bibnamefont {Korotkov}},\ and\ \bibinfo {author} {\bibfnamefont {K.~K.}\ \bibnamefont {Likharev}},\ }\bibfield  {title} {\bibinfo {title} {Theory of single-electron charging of quantum wells and dots},\ }\href {https://doi.org/10.1103/PhysRevB.44.6199} {\bibfield  {journal} {\bibinfo  {journal} {Phys. Rev. B}\ }\textbf {\bibinfo {volume} {44}},\ \bibinfo {pages} {6199} (\bibinfo {year} {1991})}\BibitemShut {NoStop}%
\bibitem [{\citenamefont {Pothier}\ \emph {et~al.}(1997)\citenamefont {Pothier}, \citenamefont {Gu\'eron}, \citenamefont {Birge}, \citenamefont {Esteve},\ and\ \citenamefont {Devoret}}]{Birge}%
  \BibitemOpen
  \bibfield  {author} {\bibinfo {author} {\bibfnamefont {H.}~\bibnamefont {Pothier}}, \bibinfo {author} {\bibfnamefont {S.}~\bibnamefont {Gu\'eron}}, \bibinfo {author} {\bibfnamefont {N.~O.}\ \bibnamefont {Birge}}, \bibinfo {author} {\bibfnamefont {D.}~\bibnamefont {Esteve}},\ and\ \bibinfo {author} {\bibfnamefont {M.~H.}\ \bibnamefont {Devoret}},\ }\bibfield  {title} {\bibinfo {title} {Energy distribution function of quasiparticles in mesoscopic wires},\ }\href {https://doi.org/10.1103/PhysRevLett.79.3490} {\bibfield  {journal} {\bibinfo  {journal} {Phys. Rev. Lett.}\ }\textbf {\bibinfo {volume} {79}},\ \bibinfo {pages} {3490} (\bibinfo {year} {1997})}\BibitemShut {NoStop}%
\bibitem [{\citenamefont {Altland}\ and\ \citenamefont {Egger}(2009)}]{AltlandEgger}%
  \BibitemOpen
  \bibfield  {author} {\bibinfo {author} {\bibfnamefont {A.}~\bibnamefont {Altland}}\ and\ \bibinfo {author} {\bibfnamefont {R.}~\bibnamefont {Egger}},\ }\bibfield  {title} {\bibinfo {title} {Nonequilibrium dephasing in coulomb blockaded quantum dots},\ }\href {https://doi.org/10.1103/PhysRevLett.102.026805} {\bibfield  {journal} {\bibinfo  {journal} {Phys. Rev. Lett.}\ }\textbf {\bibinfo {volume} {102}},\ \bibinfo {pages} {026805} (\bibinfo {year} {2009})}\BibitemShut {NoStop}%
\bibitem [{\citenamefont {Smirnov}\ and\ \citenamefont {Grifoni}(2011)}]{SmirnovKondoQD}%
  \BibitemOpen
  \bibfield  {author} {\bibinfo {author} {\bibfnamefont {S.}~\bibnamefont {Smirnov}}\ and\ \bibinfo {author} {\bibfnamefont {M.}~\bibnamefont {Grifoni}},\ }\bibfield  {title} {\bibinfo {title} {Slave-boson keldysh field theory for the kondo effect in quantum dots},\ }\href {https://doi.org/10.1103/PhysRevB.84.125303} {\bibfield  {journal} {\bibinfo  {journal} {Phys. Rev. B}\ }\textbf {\bibinfo {volume} {84}},\ \bibinfo {pages} {125303} (\bibinfo {year} {2011})}\BibitemShut {NoStop}%
\bibitem [{\citenamefont {Kouwenhoven}\ \emph {et~al.}(1991)\citenamefont {Kouwenhoven}, \citenamefont {van~der Vaart}, \citenamefont {Johnson}, \citenamefont {Kool}, \citenamefont {Harmans}, \citenamefont {Williamson}, \citenamefont {Staring},\ and\ \citenamefont {Foxon}}]{StaircaseExpt}%
  \BibitemOpen
  \bibfield  {author} {\bibinfo {author} {\bibfnamefont {L.~P.}\ \bibnamefont {Kouwenhoven}}, \bibinfo {author} {\bibfnamefont {N.~C.}\ \bibnamefont {van~der Vaart}}, \bibinfo {author} {\bibfnamefont {A.~T.}\ \bibnamefont {Johnson}}, \bibinfo {author} {\bibfnamefont {W.}~\bibnamefont {Kool}}, \bibinfo {author} {\bibfnamefont {C.~J. P.~M.}\ \bibnamefont {Harmans}}, \bibinfo {author} {\bibfnamefont {J.~G.}\ \bibnamefont {Williamson}}, \bibinfo {author} {\bibfnamefont {A.~A.~M.}\ \bibnamefont {Staring}},\ and\ \bibinfo {author} {\bibfnamefont {C.~T.}\ \bibnamefont {Foxon}},\ }\bibfield  {title} {\bibinfo {title} {Single electron charging effects in semiconductor quantum dots},\ }\href {https://doi.org/10.1007/BF01307632} {\bibfield  {journal} {\bibinfo  {journal} {Z. Phys. B Con. Mat.}\ }\textbf {\bibinfo {volume} {85}},\ \bibinfo {pages} {367} (\bibinfo {year} {1991})}\BibitemShut {NoStop}%
\bibitem [{\citenamefont {Jauho}\ \emph {et~al.}(1994)\citenamefont {Jauho}, \citenamefont {Wingreen},\ and\ \citenamefont {Meir}}]{MeirWingreenJauho}%
  \BibitemOpen
  \bibfield  {author} {\bibinfo {author} {\bibfnamefont {A.-P.}\ \bibnamefont {Jauho}}, \bibinfo {author} {\bibfnamefont {N.~S.}\ \bibnamefont {Wingreen}},\ and\ \bibinfo {author} {\bibfnamefont {Y.}~\bibnamefont {Meir}},\ }\bibfield  {title} {\bibinfo {title} {Time-dependent transport in interacting and noninteracting resonant-tunneling systems},\ }\href {https://doi.org/10.1103/PhysRevB.50.5528} {\bibfield  {journal} {\bibinfo  {journal} {Phys. Rev. B}\ }\textbf {\bibinfo {volume} {50}},\ \bibinfo {pages} {5528} (\bibinfo {year} {1994})}\BibitemShut {NoStop}%
\bibitem [{\citenamefont {Rammer}\ and\ \citenamefont {Smith}(1986)}]{RammerSmith}%
  \BibitemOpen
  \bibfield  {author} {\bibinfo {author} {\bibfnamefont {J.}~\bibnamefont {Rammer}}\ and\ \bibinfo {author} {\bibfnamefont {H.}~\bibnamefont {Smith}},\ }\bibfield  {title} {\bibinfo {title} {Quantum field-theoretical methods in transport theory of metals},\ }\href {https://doi.org/10.1103/RevModPhys.58.323} {\bibfield  {journal} {\bibinfo  {journal} {Rev. Mod. Phys.}\ }\textbf {\bibinfo {volume} {58}},\ \bibinfo {pages} {323} (\bibinfo {year} {1986})}\BibitemShut {NoStop}%
\bibitem [{\citenamefont {Haug}\ and\ \citenamefont {Jauho}(1998)}]{HaugJauho}%
  \BibitemOpen
  \bibfield  {author} {\bibinfo {author} {\bibfnamefont {H.}~\bibnamefont {Haug}}\ and\ \bibinfo {author} {\bibfnamefont {A.-P.}\ \bibnamefont {Jauho}},\ }\href@noop {} {\emph {\bibinfo {title} {Quantum kinetics in transport and optics of semiconductors}}}\ (\bibinfo  {publisher} {Springer},\ \bibinfo {address} {Berlin},\ \bibinfo {year} {1998})\BibitemShut {NoStop}%
\bibitem [{\citenamefont {McArdle}\ \emph {et~al.}(2023)\citenamefont {McArdle}, \citenamefont {Davies}, \citenamefont {Lerner},\ and\ \citenamefont {Yurkevich}}]{AsymCBPaper}%
  \BibitemOpen
  \bibfield  {author} {\bibinfo {author} {\bibfnamefont {G.}~\bibnamefont {McArdle}}, \bibinfo {author} {\bibfnamefont {R.}~\bibnamefont {Davies}}, \bibinfo {author} {\bibfnamefont {I.~V.}\ \bibnamefont {Lerner}},\ and\ \bibinfo {author} {\bibfnamefont {I.~V.}\ \bibnamefont {Yurkevich}},\ }\bibfield  {title} {\bibinfo {title} {Coulomb staircase in an asymmetrically coupled quantum dot},\ }\href@noop {} {\bibfield  {journal} {\bibinfo  {journal} {Journal of Physics: Condensed Matter}\ }\textbf {\bibinfo {volume} {35}},\ \bibinfo {pages} {475302} (\bibinfo {year} {2023})}\BibitemShut {NoStop}%
\bibitem [{\citenamefont {Agam}\ \emph {et~al.}(1997)\citenamefont {Agam}, \citenamefont {Wingreen}, \citenamefont {Altshuler}, \citenamefont {Ralph},\ and\ \citenamefont {Tinkham}}]{AltshulerTinkham}%
  \BibitemOpen
  \bibfield  {author} {\bibinfo {author} {\bibfnamefont {O.}~\bibnamefont {Agam}}, \bibinfo {author} {\bibfnamefont {N.}~\bibnamefont {Wingreen}}, \bibinfo {author} {\bibfnamefont {B.}~\bibnamefont {Altshuler}}, \bibinfo {author} {\bibfnamefont {D.}~\bibnamefont {Ralph}},\ and\ \bibinfo {author} {\bibfnamefont {M.}~\bibnamefont {Tinkham}},\ }\bibfield  {title} {\bibinfo {title} {Chaos, interactions, and nonequilibrium effects in the tunneling resonance spectra of ultrasmall metallic particles},\ }\href {https://doi.org/10.1103/PhysRevLett.78.1956} {\bibfield  {journal} {\bibinfo  {journal} {Phys. Rev. Lett.}\ }\textbf {\bibinfo {volume} {78}},\ \bibinfo {pages} {1956} (\bibinfo {year} {1997})}\BibitemShut {NoStop}%
\end{thebibliography}%


\begin{thebibliography}{25}%
\makeatletter
\providecommand \@ifxundefined [1]{%
 \@ifx{#1\undefined}
}%
\providecommand \@ifnum [1]{%
 \ifnum #1\expandafter \@firstoftwo
 \else \expandafter \@secondoftwo
 \fi
}%
\providecommand \@ifx [1]{%
 \ifx #1\expandafter \@firstoftwo
 \else \expandafter \@secondoftwo
 \fi
}%
\providecommand \natexlab [1]{#1}%
\providecommand \enquote  [1]{``#1''}%
\providecommand \bibnamefont  [1]{#1}%
\providecommand \bibfnamefont [1]{#1}%
\providecommand \citenamefont [1]{#1}%
\providecommand \href@noop [0]{\@secondoftwo}%
\providecommand \href [0]{\begingroup \@sanitize@url \@href}%
\providecommand \@href[1]{\@@startlink{#1}\@@href}%
\providecommand \@@href[1]{\endgroup#1\@@endlink}%
\providecommand \@sanitize@url [0]{\catcode `\\12\catcode `\$12\catcode `\&12\catcode `\#12\catcode `\^12\catcode `\_12\catcode `\%12\relax}%
\providecommand \@@startlink[1]{}%
\providecommand \@@endlink[0]{}%
\providecommand \url  [0]{\begingroup\@sanitize@url \@url }%
\providecommand \@url [1]{\endgroup\@href {#1}{\urlprefix }}%
\providecommand \urlprefix  [0]{URL }%
\providecommand \Eprint [0]{\href }%
\providecommand \doibase [0]{https://doi.org/}%
\providecommand \selectlanguage [0]{\@gobble}%
\providecommand \bibinfo  [0]{\@secondoftwo}%
\providecommand \bibfield  [0]{\@secondoftwo}%
\providecommand \translation [1]{[#1]}%
\providecommand \BibitemOpen [0]{}%
\providecommand \bibitemStop [0]{}%
\providecommand \bibitemNoStop [0]{.\EOS\space}%
\providecommand \EOS [0]{\spacefactor3000\relax}%
\providecommand \BibitemShut  [1]{\csname bibitem#1\endcsname}%
\let\auto@bib@innerbib\@empty

\bibitem{MeirWingreenJauhoSupp}
A.-P. Jauho, N.~S. Wingreen, and Y. Meir, {\em Phys. Rev. B} {\bf 50},  5528  (1994).

\bibitem{RammerSmithSupp}
J. Rammer and H. Smith, {\em Rev. Mod. Phys.} {\bf 58},  323  (1986).

\bibitem{HaugJauhoSupp}
H. Haug and A.-P. Jauho, {\em Quantum kinetics in transport and optics of semiconductors} (Springer, Berlin, 1998).

\bibitem{AsymCBPaperSupp}
G. McArdle, R. Davies, I.~V. Lerner, and I.~V. Yurkevich, {\em Journal of Physics: Condensed Matter} {\bf 35},  475302  (2023).

\bibitem{BeenakkerSupp}
C.~W.~J. Beenakker, {\em Phys. Rev. B} {\bf 44},  1646  (1991).

\bibitem{AverinLikharevBookChapterSupp}
D. Averin and K. Likharev,  in {\em Mesoscopic Phenomena in Solids}, Vol.~30 of {\em Modern Problems in Condensed Matter Sciences}, edited by B. Altshuler, P. Lee, and R. Webb (Elsevier, Amsterdam, 1991), pp.\ 173--271.

\bibitem{KulikSupp}
I.~O. Kulik and R.~I. Shekhter, {\em Zh. Eksp. Teor. Fiz.} {\bf 68},  623  (1975).

\bibitem{Ben-JacobWilkinsSupp}
M. Amman {\it et~al.}, {\em Phys. Rev. B} {\bf 43},  1146  (1991).

\bibitem{AvLikKorSupp}
D.~V. Averin, A.~N. Korotkov, and K.~K. Likharev, {\em Phys. Rev. B} {\bf 44},  6199  (1991).

\end{thebibliography}

\clearpage

\setcounter{equation}{0}
\setcounter{figure}{0}
\setcounter{table}{0}
\setcounter{page}{1}
\makeatletter

\renewcommand{\theequation}{S\arabic{equation}}
\renewcommand{\thefigure}{S\arabic{figure}}
\renewcommand{\bibnumfmt}[1]{[S#1]}
\renewcommand{\citenumfont}[1]{S#1}

\onecolumngrid

\begin{center}
\large{\textbf{Supplemental material for Coulomb blockade in a {non-thermalized} quantum dot}}
\end{center}

{\section{Green's functions for an isolated quantum dot and the incorporation of tunneling to the leads}

The Green's function for an isolated dot can be found explicitly, as we demonstrate below and then in the limit of weak coupling to the leads, the quantum kinetic equation can be derived using the non-equilibrium Keldysh formalism (see, e.g.,  \cite{MeirWingreenJauhoSupp, RammerSmithSupp, HaugJauhoSupp}), allowing the Green's function of the dot coupled to leads to be found.
In the limit where the broadening of the levels due to tunneling to and from the leads, $\Gamma$, is much less than the mean level spacing, $\Delta$, then the Green's functions of a dot, $g(\varepsilon)$ can be expressed as a sum over the Green's functions for individual energy levels, $n$, so that $g(\varepsilon)=\sum_n g_n(\varepsilon)$. The greater and lesser functions are given by $g^>_n(t) = -i \langle d_n(t)d_n^\dagger(0) \rangle$ and $g^<_n(t) = i \langle d_n^\dagger(0)  d_n(t) \rangle$ respectively, where $d_n(t) = \mathrm{e}^{iHt}d_n(0) \mathrm{e}^{-iHt}$ annihilates an electron on the dot. In the case of an isolated dot, the Hamiltonian is given by Eq.~(4a) of the main text, which ensures that the number of electrons on the dot, $N$, is conserved allowing the Green's functions to be expressed as sums over $N$,
\begin{align}
    g^>_n(\varepsilon) &= - i \sum_N \mathrm{Tr}_N \left(\hat{\rho}_0 \mathrm{e}^{iHt}d_n \mathrm{e}^{-iHt}d_n^\dagger(0)\right), \\
    g^<_n(\varepsilon) &=  i \sum_N \mathrm{Tr}_N \left(\hat{\rho}_0 d_n^\dagger(0)\mathrm{e}^{iHt}d_n \mathrm{e}^{-iHt}\right).
\end{align}
Here $\hat{\rho}_0$ is the density matrix for an isolated dot and $\mathrm{Tr}_N$ is the trace in the subspace where there are $N$ electrons on the dot. Explicit evaluation of these leads to, after Fourier transforming, the Green's functions of the isolated dot \cite{AsymCBPaperSupp}
\begin{eqnarray}
	\label{g>_isolated}
	g^>_n(\varepsilon) = -2\pi i \sum_N \delta \left(\varepsilon - \varepsilon_n - \Omega_N \right) g^>_N(\varepsilon_n), \hspace{10pt} &g^>_N(\varepsilon_n) = \Tr_N \left(\hat{\rho}_0 d_n d_n^\dagger \right), \\
	\label{g<_isolated}
    g^<_n(\varepsilon) = -2\pi i \sum_N \delta \left(\varepsilon - \varepsilon_n - \Omega_{N-1} \right) g^<_N(\varepsilon_n), \hspace{10pt} &g^<_N(\varepsilon_n) = -\Tr_N \left(\hat{\rho}_0 d_n^\dagger d_n \right).
\end{eqnarray}
By defining the probability of having $N$ electrons on the dot, $p_N$, and the associated distribution function for this number of electrons, $F_N(\varepsilon_n)$, we use the ansatz $g^>_N(\varepsilon_n) = p_N\left(1-F_N(\varepsilon_n)\right) \hspace{2pt} \mathrm{ and} \hspace{5pt} g^<_N(\varepsilon_n) = -p_N F_N(\varepsilon_n)$. This leads to the results in Eq.~(8) of the main text. In order to incorporate the tunneling to the leads, the standard quantum kinetic equation is used for each level on the dot \cite{MeirWingreenJauhoSupp,RammerSmithSupp,HaugJauhoSupp}
\begin{equation}\label{QKE_original}
	g^{>,<}_n \left(\varepsilon\right) = g^{\mathrm{R}}_n \left(\varepsilon\right) \Sigma^{>,<} \left(\varepsilon\right) g^{\mathrm{A}}_n \left(\varepsilon\right).
\end{equation}
where we have used that the full Green's function of the dot can be replaced by that of the isolated dot in the weak coupling limit, $\Gamma \rightarrow 0$. Combining the two equations in Eq.~(\ref{QKE_original}) leads to the form of the kinetic equation given in Eq.~(7) of the main text,
\begin{equation}
    g_n^>(\varepsilon)\Sigma^<(\varepsilon) = g_n^<(\varepsilon) \Sigma^>(\varepsilon).
\end{equation}
Upon substitution of the forms of the Green's functions and self-energies laid out in the main text (Eq.(8) and (9) respectively), the QKE has the form
 \begin{align}\label{QKE}
    p_{N+1}F_{N+1}(\varepsilon_n)&\left(1-\widetilde{f}(\varepsilon_n + \Omega_N) \right) = p_N \left(1-F_N(\varepsilon_n)\right)\widetilde{f}(\varepsilon_n + \Omega_N),\\
\widetilde{f}(\varepsilon)& = (\Gamma _{\mathrm{L}}/\Gamma)f_{\mathrm{L}}(\varepsilon) + (\Gamma _{\mathrm{R}}/\Gamma)f_{\mathrm{R}}(\varepsilon) .\notag
\end{align}
similar to the detailed balance relations derived in \cite{BeenakkerSupp}.

The results of the QKE, that is $p_N$ and $F_N(\varepsilon)$ can then be subsequently used to calculate the current. To see this, consider the standard expression for tunneling current going from the dot to the lead, $\alpha$ \cite{MeirWingreenJauhoSupp}
\begin{equation}
	I_\alpha = -ie\Gamma_\alpha \int \frac{\mathrm{d}\varepsilon}{2\pi} \left(g^<(\varepsilon)+f_\alpha(\varepsilon)\left[g^>(\varepsilon)-g^<(\varepsilon) \right] \right).
\end{equation}
After substitution of the dot Green's functions in Eq.~(\ref{g>_isolated}, \ref{g<_isolated}) along with the corresponding ansatz, into this formula for the current, then the current can be expressed as
\begin{equation}
	I_\alpha = e\Gamma_\alpha \sum_N p_N \sum_n \bigg(F_N(\varepsilon_n)[1-f_\alpha(\varepsilon_n+\Omega_{N-1})] - [1-F_N(\varepsilon_n)]f_\alpha(\varepsilon_n+\Omega_N) \bigg).
\end{equation}
By utilising current conservation, the current can then be expressed as
\begin{equation}\label{Current}
	I = e\frac{\Gamma_\mathrm{L} \Gamma_\mathrm{R}}{\Gamma} \sum_{N,n} p_N \Big( F_N(\varepsilon_n)\left[f_\mathrm{L}(\varepsilon_n +\Omega_{N-1}) - f_\mathrm{R}(\varepsilon_n +\Omega_{N-1})\right] + (1-F_N(\varepsilon_n)) \left[f_\mathrm{L}(\varepsilon_n +\Omega_N) - f_\mathrm{R}(\varepsilon_n +\Omega_N) \right] \Big),
\end{equation}
as is presented in Eq.(5) of the main text. We have calculated the current in the case when only two consecutive occupation numbers, say $N$ and $N+1$, contribute, but the above equation is applicable more generally. Outside the Coulomb blockade regime, specified by inequalities (3), one just needs to modify Eq.(6) for $\Omega_N$.}

\section{Full solution to the quantum kinetic equation}

In order to calculate the current through the quantum dot using Eq.~(5) of the main text, it is necessary to find the probability that it has $N$ electrons, $p_N$, and the distribution function given that it has $N$ electrons, $F_N(\varepsilon)$. To do this, we make use of the quantum kinetic equation (QKE) which for a quantum dot in the Coulomb blockade regime coupled to two leads is given by Eq.~(10) in the main text
\begin{equation}\label{QKE}
	p_N \left(1-F_N(\varepsilon_n)\right)\tilde{f}(\varepsilon_n + \Omega_N) = p_{N+1}F_{N+1}(\varepsilon_n)\left(1-\tilde{f}(\varepsilon_n + \Omega_N) \right).
\end{equation}
In this equation, $\tilde{f}(\varepsilon) = \frac{\Gamma_{\mathrm{L}}}{\Gamma}f_{\mathrm{L}}(\varepsilon) + \frac{\Gamma_{\mathrm{R}}}{\Gamma}f_{\mathrm{R}}(\varepsilon)$ and the absence of thermalization on the dot has been assumed.
Providing there are only two states whose probabilities aren't exponentially suppressed this has a solution in which $F_N(\varepsilon_n) \approx F_{N+1}(\varepsilon_n) \approx F(\varepsilon_n)$. The probabilities are then found from the normalization conditions,
 \begin{equation}\label{Normalisation}
 	\int_0^\infty F_N(\varepsilon) \mathrm{d}\varepsilon = N\Delta \equiv \varepsilon_\mathrm{F}, \quad \sum_N p_N = 1,
 \end{equation}
 where the energies in the dot are counted from the bottom of the band and the normalization of the probabilities can be written as $p_N + p_{N+1} \approx 1$. This solution is valid in the limit $N \gg 1$ and here we demonstrate how this solution is obtained using the saddle point approximation as we achieved in \cite{AsymCBPaperSupp}. When there are only two relevant probabilities, Eq.~(\ref{QKE}) has an exact solution
 \begin{alignat}{2}
 	p_N &= \frac{Z_N}{Z_N+Z_{N+1}}, &p_{N+1} &= \frac{Z_{N+1}}{Z_N+Z_{N+1}}, \nonumber \\ \label{Full_Soln} \\
 	F_N(\varepsilon_n) &= \frac{Z_N(\varepsilon_n)}{Z_N}, &F_{N+1}(\varepsilon_n) &= \frac{Z_{N+1}(\varepsilon_n)}{Z_{N+1}}. \nonumber
 \end{alignat}
 Introducing
 \begin{equation}\label{varphi}
	\varphi(\varepsilon) = \frac{\tilde{f}(\varepsilon)}{1 - \tilde{f}(\varepsilon)},
\end{equation}
 $Z_N$ and $Z_{N+1}$ are defined as
 \begin{align}
 	Z_N &= \sum_{\{n_j=0,1\}} \prod_{j=1}^\infty \left[\varphi(\varepsilon_j+\Omega_N)\right]^{n_j} \delta_{(\sum_j n_j), N},
\nonumber \\ \label{Z_defn} \\ \nonumber
 	Z_{N+1} &= \sum_{\{n_j=0,1\}}  \prod_{j=1}^\infty \left[\varphi(\varepsilon_j+\Omega_N)\right]^{n_j} \delta_{(\sum_j n_j), N+1}.
 \end{align}
 Then $Z_N(\varepsilon_n)$ and $Z_{N+1}(\varepsilon_n)$, required for calculating the distribution functions in Eq.~(\ref{Full_Soln}), are found by restricting the sums  in Eq.~(\ref{Z_defn}) to terms where the level $\varepsilon_n$ is occupied. We stress that in these definitions the relevant $N$ dependence enters only via the    Kr\"{o}necker delta's as both $Z_N$ and $Z_{N+1}$ contain $\varphi(\varepsilon_j+\Omega_N)$, reflecting the fact that the two states are coupled via the QKE, Eq.~(\ref{QKE}), which contains $\tilde{f}(\varepsilon_n + \Omega_N)$.
 {The Kr\"{o}necker delta can be written as an integral}
 \begin{equation}\label{Eq:Delta_n_sum}
	\delta_{(\sum_j n_j), N} = \int \frac{\mathrm{d}\theta}{2\pi}\mathrm{e}^{i\theta\left(\sum_j n_j - N\right)},
\end{equation}
meaning that $Z_N$ can be written in a form which we evaluate using the saddle-point approximation.
\begin{equation}\label{Z_saddle}
	Z_N = \int \frac{\mathrm{d}\theta}{2\pi} \mathrm{e}^{Nf(\theta)}, \qquad f(\theta) = \frac{1}{N}\sum_j \ln\left( 1+ \varphi(\varepsilon_j+\Omega_N)\mathrm{e}^{i\theta}\right) -i\theta.
\end{equation}
Recalling that the density of states in the dot is approximately the inverse of the mean level spacing, $\Delta^{-1}$, we write the sum in the definition of $f(\theta)$ as an integral, so that the saddle point, $\theta_0$, is determined by the equation
\begin{equation}\label{Saddle_Point}
\varepsilon_\mathrm{F} =
\int_0^\infty {\mathrm{d}\varepsilon} \frac{\varphi(\varepsilon + \Omega_N)}{\varphi(\varepsilon + \Omega_N)+\mathrm{e}^{-i\theta_0}}.
\end{equation}
Despite the presence of $\Omega_N$, the relevant $N$ dependence of $\theta_0$ enters only via $\varepsilon_\mathrm{F} \equiv N\Delta$, as there is no change in $\Omega_N$ going from $Z_N$ to $Z_{N+1}$. Therefore we write $Z_N = g(\theta_0)\mathrm{e}^{-iN\theta_0}$, where $g(\theta_0)$ is a function depending on $N$ only through $\varepsilon_\mathrm{F}$. In the limit $N \gg 1$, $N\Delta \approx (N+1)\Delta$, so that $\varepsilon_\mathrm{F}$ is approximately a constant and consequently $g(\theta_0)$ is approximately the same for $Z_N$ and $Z_{N+1}$. Therefore we find, after using Eq.~(\ref{Full_Soln}), that
\begin{equation}\label{p and F}
	\frac{p_{N+1}}{p_N} = \mathrm{e}^{-i\theta_0}, \hspace{10pt} F_N(\varepsilon_n)   \approx F_{N+1}(\varepsilon_n) \approx \left({\frac{\mathrm{e}^{-i\theta_0}}{\varphi(\varepsilon + \Omega_N)}+1}\right)^{-1}.
\end{equation}
This solution is equivalent to assuming $F_N(\varepsilon_n) \approx F_{N+1}(\varepsilon_n)$ in the QKE, Eq.~(\ref{QKE}), with the ratio of probabilities then being given by the normalization of $F_N(\varepsilon_n)$, Eq.~(\ref{Normalisation}) (or equivalently Eq.~(\ref{Saddle_Point})). This is the solution presented in the main text.

{\section{Normalization of the non-thermalized distribution function}
In  {the regime characterized  by the inequalities (1) and (3) of the main text}, $\widetilde{f}(\varepsilon_ n+\Omega_N) = (\Gamma_\mathrm{L}/\Gamma)f_\mathrm{L}(\varepsilon_ n+\Omega_N)+ (\Gamma_\mathrm{R}/\Gamma)f_\mathrm{R}(\varepsilon_ n+\Omega_N)$ can be split into three regions,
\begin{equation}\label{f tilde regions}
	\widetilde {f}(\varepsilon_n +\Omega_N) \approx \begin{cases}
		1 -(\Gamma _{\mathrm{R}}/\Gamma) \mathrm{e}^{\beta(\varepsilon_n-(\mu-\Omega_N))}, &\varepsilon_n < \mu-\Omega_N \\
		\Gamma _{\mathrm{L}}/\Gamma,  &\mu-\Omega_N < \varepsilon_n < \mu-\Omega_N+eV \\
		(\Gamma _{\mathrm{L}}/\Gamma) \mathrm{e}^{-\beta[\varepsilon_n-(\mu -\Omega_N+eV)]}, & \mu -\Omega_N+eV<\varepsilon_n.
	\end{cases}
\end{equation}
Using the normalization of $F(\varepsilon_n)$ with $\varepsilon_\mathrm{F} = N\Delta$
\begin{equation}
	\varepsilon_\mathrm{F} = \int_0^\infty \mathrm{d}\varepsilon F(\varepsilon_n) = \int_0^\infty \mathrm{d}\varepsilon \frac{\widetilde{f}(\varepsilon + \Omega_N)}{[1-\widetilde{f}(\varepsilon + \Omega_N)]A_N+\widetilde{f}(\varepsilon + \Omega_N)}
\end{equation}
and making use of the three regions in Eq.~(\ref{f tilde regions}) results in the following equation to determine $A_N$:
\begin{align}\label{AN Eqn}
	\beta \varepsilon_\mathrm{F} &= \frac{\beta e V}{A_N\frac{\Gamma_\mathrm{R}}{\Gamma_\mathrm{L}}+1} +\ln \left( \frac{\Gamma}{\Gamma _{\mathrm{R}}A_N}\mathrm{e}^{\beta(\mu - \Omega_N)}+1\right)  + \ln \left(\frac{\frac{\Gamma _{\mathrm{L}}}\Gamma  +   A_N } {\frac\Gamma{\Gamma _{\mathrm{R}}}  +   A_N }   \right).
\end{align}
This equation can be solved numerically across the entire voltage range, $0 \leq eV \lesssim \Omega_{N+1}$   leading to the probabilities $p_N$ and $p_{N+1}$ as shown in Fig.~1 of the main text.
In the vicinity of the additional jump in the differential conductance, where $A_N \sim 1$, a useful analytical estimate for $A_N$ can be obtained, which allows for an estimation of the size of the jump not present in the thermalized case \cite{AverinLikharevBookChapterSupp, KulikSupp, Ben-JacobWilkinsSupp, AvLikKorSupp}. Since $\mu = \varepsilon_\mathrm{F} \gg E_\mathrm{c}$, then the first logarithm in Eq.~(\ref{AN Eqn}) can be simplified to give
\begin{align}
	\beta \varepsilon_\mathrm{F} &\approx \frac{\beta e V}{A_N\frac{\Gamma_\mathrm{R}}{\Gamma_\mathrm{L}}+1} + \beta(\mu-\Omega_N) +  \ln \left( \frac{\Gamma}{\Gamma _{\mathrm{R}}A_N}\right)  + \ln \left(\frac{\frac{\Gamma _{\mathrm{L}}}\Gamma  +   A_N } {\frac\Gamma{\Gamma _{\mathrm{R}}}  +   A_N }   \right).
\end{align}
The first term on the right-hand side of this equation is of the order $\beta E_\mathrm{c} \gg 1$ near the additional jump in conductance and as the arguments of the logarithms are $\mathcal{O}(1)$, then it can be found that close to the jump
\begin{equation}\label{Jump AN}
	A_N \approx \frac{\Gamma_\mathrm{L}}{\Gamma_\mathrm{R}}\left(\frac{eV-\Omega_N}{\Omega_N} \right)
\end{equation}
as presented in the main text. By evaluating the current (Eq.~(13) of the main text) on either side of the jump, which occurs at $eV=E_\mathrm{c}$, the size of the jump in the low temperature limit is found to be,
\begin{equation}
	\delta G =  \frac{e^2}{\Delta}\frac{\Gamma_\mathrm{L}\Gamma_\mathrm{R}}{\Gamma}\frac{1+\tfrac{\Gamma_\mathrm{R}}{\Gamma_\mathrm{L}}\tilde{A}_N^2}{(1+\tilde{A}_N)(1+\tfrac{\Gamma_\mathrm{R}}{\Gamma_\mathrm{L}}\tilde{A}_N)}
\end{equation}
with $\tilde{A}_N$ being given by Eq.~(\ref{Jump AN}) evaluated at the position of the jump, $eV=E_\mathrm{c}$. This leads to the result of Eq.~(15) in the main text for $\Omega_N = E_\mathrm{c}/2$.
}

\end{document}